%% file: FakeNewsAfterReviews.tex
\documentclass[letterpaper, 10 pt, conference]{ieeeconf}

\usepackage{ulem}
\usepackage{multirow}

\usepackage{enumerate}

\usepackage{booktabs} 
\usepackage{tikz}
\usepackage[theorems,skins]{tcolorbox}
\usepackage{etoolbox}
\usepackage{amsmath}
\usepackage{caption}
\usepackage{physics}
\usepackage{hyperref}

\usepackage{graphicx}
\usepackage{siunitx}
\usepackage{import}
\usepackage{placeins}

\usepackage{amsthm}
\usepackage{amssymb}
\usepackage{empheq}
\usepackage{ulem}
\usepackage{graphicx}
\usepackage[noend]{algpseudocode}
\usepackage{algorithm}
\usepackage{placeins}

\newcommand{\ignoreDistinguish}[1]{}
\newcommand{\resp}[1]{{\color{blue}#1}}
\newcommand{\eop}{\hfill{$\blacksquare$}}

\newtheorem{thm}{Theorem}

\newtheorem{lem}{Lemma}

\newcommand{\N}{\mathcal{N}}
\usepackage{bm}
\usepackage{framed}

\usepackage{ifthen}
\usepackage{float}
\newcommand{\Detail}[1]{}

 \newcommand{\ignore}[1]{}

\usepackage{multicol}
 \usepackage{caption}
\usepackage{subcaption}
\newcommand{\Arxivtwo}[2]{#1}  

\begin{document}
\Arxivtwo{\title{Controlling Fake News by  Tagging: A Branching Process Analysis}}
{\title{Controlling Fake News by  Collective Tagging: A Branching Process Analysis}}

  \author{Suyog Kapsikar$^1$, Indrajit Saha$^1$, Khushboo Agarwal$^1$, Veeraruna Kavitha$^1$, and  Quanyan Zhu$^2$    \\
   $^1$IEOR, IIT Bombay, India, $^2$New York University, USA
   }
\maketitle
\thispagestyle{empty} 
\begin{abstract}
The spread of fake news on online social networks (OSNs) has become a matter of concern. These platforms are also used for propagating important authentic information. Thus, there is a need for mitigating fake news without significantly influencing the spread of real news. We leverage users' inherent capabilities of identifying fake news and propose a warning-based control mechanism to curb this spread. Warnings are  based on previous users' responses  that indicate the authenticity of the news. We use population-size dependent continuous-time multi-type  branching processes to describe the spreading under the warning mechanism. We  also have new results towards these  branching processes. The (time) asymptotic proportions of  the individual populations are derived using stochastic approximation tools.
Using these, relevant type $1$, type $2$ performances are derived and 
an appropriate optimization problem is solved. The proposed mechanism effectively controls fake news, with negligible influence on the propagation of authentic news. We validate performance measures using Monte Carlo simulations on  network connections provided by  Twitter data.
\end{abstract}
\section{Introduction}
\vspace{-1mm}
Fake news is fabricated (mis)information that propagates through social  media like authentic news \cite{lazer2018science}. It does not go through the same scrutiny as the news from the legit news media source. Fake news  has varying degrees of impact on users and society. It can influence the political choices of the users (e.g., \cite{Allcott} discusses fake news articles during the U.S. 2016 elections), can impact financial stock markets (e.g., \cite{Kogan}), etc. Thus, it is essential to address growing concerns of fake news and develop an intervention policy.

Based on the empirical database, it has been found that fake news propagates differently from real news on OSNs. In  \cite{Vosoughi1146}, the authors have shown that fake news propagates faster and farther.  In \cite{zhao2018fake}, the authors discuss the differences between the propagations using the evidence based on empirical data of (fake and real) posts from Twitter Japan. 
The above discussions demonstrate a few important facts: a) users could be more attracted to fake news \cite{Vosoughi1146}; b) there are differences between fake   and real news propagations. We assume, users can use their cognitive judgement and reasoning capabilities to differentiate between fake and real news items to some extent. It may not yield  perfect detection of fake news, but  \textit{appropriate aggregation of such judgements (a.k.a. collective wisdom) can probably deter the fake news spreading.}

Branching processes (BPs) are  used  to model several problems related to content propagation over OSNs,  e.g.,  meme popularity (analysis of general posts)  over multiple networks in (\cite{d2019spreading, yagan2013conjoining}), viral marketing (propagation of a particular  post of interest  as in our case) in (\cite{viralBranching,ranbir}). In OSNs  (e.g., Facebook), users share posts with their friends. When a friend visits\footnote{visits OSN, opens his timeline and reads the news/post.} the OSN,  he may forward   the post to some of his friends, depending on   attractiveness.  This spreading is sufficiently well captured by continuous-time Markovian (exponential user visit times) BPs, which are also mathematically tractable (e.g., \cite{viralBranching,ranbir}). Continuous versions can model independent user visits   as well as the possible bursty spread (large   shares in short time-intervals) of   post. 

While mitigating the spread of fake news, the impact on the propagation of authentic news should be minimized. To this end, we propose a new warning-based mechanism, in which every user: a)  receives a warning for each news item   he receives; b)  receives a tag from its sender indicating   the news is fake/real; and c)  tags the news as fake/real before sharing to his friends,  and the number of shares should not depend upon the tag. 
This  approach is based on {\it two assumptions}: (i) {the users have an innate capacity (see \cite{zhou2018fake}) to identify the fake news (to some extent), which can be significantly accentuated by well-designed warnings, (ii) the collective wisdom (possible by sharing) (as in \cite{landemore2012collective})  guided by `good' warnings can lead to almost unanimous and correct tagging}. The design of warning is based on the judgment (tags) of   previous users and  on the network's prior knowledge   about the  news.  Some users may not follow the rules,  may be reluctant to share  a post after tagging it as fake. We study the affect of users' reluctance on our mechanism in \cite{arxiv}.

The warning-controlled content propagation can be modelled (only) using   population size-dependent continuous-time multi-type BPs. These specialised BPs have been studied to a relatively smaller extent (e.g., \cite{gonzalez2004multitype}  considers a discrete-time version). To the best of our knowledge, the continuous-time multi-type versions of BPs have not been considered in this context.
\textit{We derive time-asymptotic proportions of individual populations} using  \textit{stochastic approximation techniques}. Such an amalgam of stochastic approximation and branching processes is not seen before.   

 We define type-$1$ and type-$2$ performance measures to quantify the impact of controlled warnings on the propagation of fake and real news. With optimal parameters of a relevant optimization problem, the type-$1$ performance improves significantly, e.g., only $10\%$ of smart users ($20\%$ of average users)
 mis-tag the fake news as real, when type-$2$ performance is within  $2\%$. In contrast, in uncontrolled system (where users tag based on system's prior knowledge and their innate capacity), $72\%$ of users mis-tag the fake news. 
 We validate the performance measures using Monte Carlo simulations on network datasets from  Twitter (\cite{snapnets}).

\vspace{-2mm}
\section{System Description}
\label{Sec_System_model}
We consider an OSN with a sufficiently large user base as in Facebook or Twitter.
The news posts on the OSN can be fake or real, and any news can be tagged as fake or real by the users. When a user with an unread copy of the news (tagged as fake or real by its sender) visits the OSN, he can view the sender's tag and a system-provided warning. Based on these two pieces of information and the user's \textit{cognitive and reasoning capability} to recognize the authenticity of the news, the user tags the news as fake or real and forwards the same to his friends. This  results in more unread copies of the news tagged as fake or real. 
This process continues, when another user (with news) visits the OSN. The warnings are designed by the system and depend on the tags of the previous users. This propagation dynamics can be captured by a population size-dependent continuous-time multi-type BP (CTMTBP).
 
\vspace{1mm}
\noindent \textbf{News Propagation and Branching Process:}  To capture the above-controlled news propagation
dynamics, we model the same using a two-type branching process $(X(t), Y(t))$. Here, $X(t)$ represents the number of users that have received the news tagged as fake but have not read/shared it yet (i.e., the number of unread copies of post with fake-tag, referred to as $x$-users); similarly $Y(t)$ represents the number of users that have received the news with real-tag (referred to as $y$-users). The dynamics depend upon the underlying news $u$,  which  can be fake (i.e., $u=F$) or real ($u=R$). The users visit their timeline independently after an exponentially distributed time with  (known) parameter $\lambda$ (as in \cite{viralBranching,ranbir}).

When any of the users that have received news $u$ with fake-tag visits the OSN (at time $t$), and if $\omega_t$ is the warning at that time, the user tags the news as fake (real) with probability $q_F^u(\omega_t)$ (respectively, $1-q_F^u(\omega_t)$) before sharing. We model $q_F^u$ as a linear function of $\omega_t$, i.e., $q_F^u(\omega_t) = \alpha_F^u \omega_t$, where \textit{$\alpha_F^u$ is the sensitivity parameter} when the underlying news $u$ is received with fake-tag. Similarly, if a  user has received the news with real-tag, he tags it as fake/real  with  probabilities $q_R^u(\omega_t)$ and  $(1-q_R^u(\omega_t))$, respectively. Here again, $q_R^u(\omega_t)= \alpha_R^u \omega_t$. \textit{The sensitivity parameters, $\alpha_F^u$ and $ \alpha_R^u$, are associated with the user's intrinsic ability to recognize the actuality, i.e., whether the news item $u$ is real or fake. }

\vspace{1mm}
\noindent \textbf{Controlled Warning:} The warnings provided by the OSN are based on the responses of the previous users. They are specific to a news item and are generated   as follows:

\vspace{-4mm}
{\small
\begin{equation} \label{eq:warning}
 \omega_t = \bigg ( \frac{wX(t)}{X(t)+bY(t)} + \epsilon \bigg) = \bigg ( \frac{w\beta(t)}{\beta(t)+b(1 - \beta(t))} + \epsilon \bigg) ,  
\end{equation}}\noindent 
where {\small $\beta(t) := X(t) / (X(t) + Y(t))$} is the relative fraction of   copies tagged as fake at time $t$; $w$ and $b$ are the control parameters. Here, $w$ takes any positive value bounded by~$1$. A smaller $b \hspace{-0.8mm}> \hspace{-0.8mm}0$ makes warnings less sensitive to $Y(t)$ (posts with real-tag), and more sensitive to  $X(t)$. A small $\epsilon \hspace{-1mm}>\hspace{-1mm}0$ captures the warning provided by the network, independent of user tags, through some fact-check mechanism.
 
\vspace{1mm}
\noindent \textbf{Tagging and Forwarding:} When a user with an unread copy of the news tagged as fake/real, reads the news (at time $\tau$),  he forwards it to some/all of his friends based on the attractiveness of the news, represented by $\eta_u$. This parameter depends on the veracity of the news ($u$ is fake or real). Let  ${\mathcal F}$ be the number of friends of a typical user of OSN and we assume  ${\cal F}$ to be i.i.d. (independent and identically distributed) across various users. He shares to $Bin(\mathcal{F},\eta_u)$ among his friends, where $Bin(\cdot, \cdot)$ is a binomial random variable. Before sharing, he tags the news as fake or real with probabilities, $q_F^u(\omega_\tau)$ and $(1 - q_F^u(\omega_\tau))$, respectively. With fake-tag, the   $x$-population gets updated; otherwise the $y$-population gets updated. These shares can be equivalently viewed as the offsprings (of various types) produced in the BP; thus the offsprings produced~by $x$-user, has   the following probability distribution (with $m_f:= E[\mathcal{F}]$):

\vspace{-4mm}
{\small \begin{eqnarray}
\label{eqn_offspring_x}
 \xi_{xx}= \xi_{xy}= Bin(\mathcal{F},\eta_u ), \mbox{  \normalsize and } E[\xi_{xx}] = m_f \eta_u := m_\eta,
\end{eqnarray}}where $\xi_{xx}, \xi_{xy}$ are  
the  fake (new users that received the news with fake-tag) and real  offsprings respectively.
Thus the evolution of the system at the transition epoch (of $x$-user wake-up), $\tau$, is summarized\footnote{We have used the fact that a user after reading the post, will rarely read or forward again; so we assume  the number of unread posts decreases by~$1$.} as follows (see \eqref{eqn_offspring_x}):
 
\vspace{-2mm}
{\small \begin{equation}\label{eqn_transition}
\begin{aligned}
X(\tau^+) & = X(\tau^-) - 1 + T_F^x \xi_{xx}, \\
Y(\tau^+) & = Y(\tau^-) + (1 - T_F^x)\xi_{xy},
\end{aligned}
\end{equation}}where $T^x_F$ is an  indicator function indicating that the $x$-user has tagged the news as fake, and $\tau^+, \tau^-$ represent the usual limits,  e.g., $X(\tau^+) := \lim _{t\downarrow \tau}X(t)$, $X(\tau^-) := \lim _{t\uparrow \tau} X(t)$. Observe that $E[T^x_F|\mathcal{G}_t] = q_F^u(\omega_t) = \alpha_F^u \omega_t$  a.s., where $\mathcal{G}_t$ is the sigma-algebra generated by $\{X(t'), Y(t'); t'\le t\}$. 
If a $y$-user, one that has received news with real-tag, visits the OSN,   he tags it as fake with (conditional) probability $q_R^u(\omega_t)$ and   system evolves  similarly: 
 
\vspace{-5mm}
{\small \begin{eqnarray}\label{eq:transition2}
Y(\tau^+) = Y(\tau^-) - 1 + (1 - T^y_F)\xi_{yy}, \hspace{19mm}\\
X(\tau^+) = X(\tau^-) + T^y_F \xi_{yx}, \mbox{ with }  \xi_{yy}= \xi_{yx}= Bin(\mathcal{F},\eta_u ). \nonumber \hspace{-10mm}
\end{eqnarray}}
Here,  $E[T^y_F|\mathcal{G}_t] = q_R^u(\omega_t) =  \alpha_R^u \omega_t$ a.s.
 
\vspace{2mm}

Under the controlled warnings, our objective is to keep most users informed of the fake news, when $u=F$. As the first step, we analyze the system for any given $w,b$.

\vspace{1mm}
\noindent \textbf{Generator Matrix:}
The   analysis of any  BP depends upon its generator matrix, computed using  the probability generating functions (PGFs) of  the  offsprings (\cite{athreya}). 
\Arxivtwo{
For any  complex vector $s = (s_x,s_y)  $, from \eqref{eqn_offspring_x}, the  
  PGFs  equal (by conditioning on ${\cal F}$ and using the PGF of  $Bin(\cdot, \cdot)$):

\vspace{-4mm}
{\small \begin{align}
f^u_x(s) & = 
E[ s_x^{\xi_{xx}} s_y^{\xi_{xy}}] =
q_F^u(\omega) k_x  + (1-q_F^u(\omega))k_y \mbox{,  \normalsize  and, } \nonumber \\
f^u_y(s) & = q_R^u(\omega) k_x  + (1-q_R^u(\omega))k_y  \mbox{, \normalsize  where, }
\label{Eqn_PGF} \\
k_j &:= E \left [ (\eta_u   s_j + 1 - \eta_u )^{\cal F} \right ] \mbox{ \normalsize for } j \in \lbrace{x,y\rbrace}.
\nonumber
\end{align}}
%
As   in \cite{athreya}, 
the generator matrix,    
$A :=(a_{i,j})_{2\times 2}$  with

\vspace{-4mm}
{\small
\begin{align*}
a_{ij} = \lambda \left ( \left . { \frac{\partial f^u_i (s)}{\partial s_j} }\right |_{s =(1,1)} - 1_{i = j} \right ), \mbox{ for } i, j \in \{x, y\}. 
\end{align*}}   
By direct computations (see
\eqref{eq:warning}, \eqref{eqn_offspring_x}, \eqref{Eqn_PGF}):}{\resp{By direct computations, the generator matrix, (see
\eqref{eq:warning}, \eqref{eqn_offspring_x} and \cite{arxiv}):}}
 
\vspace{-4mm}
{\small{
\begin{eqnarray}\label{Eqn_Gen_matrix}
A  =  A(Z) =
\lambda
\begin{bmatrix}
 q_F^u(\omega(\beta)) m_\eta  -1 & \hspace{-2mm} \left(1-q_F^u(\omega(\beta)\right) m_\eta\\ 
q_R^u(\omega(\beta)) m_\eta  & \hspace{-2mm} \left(1-q_R^u(\omega(\beta))\right) m_\eta -1
\end{bmatrix}.\hspace{-3mm}
\end{eqnarray}}}Observe that $A$   depends on  population sizes $Z = (X, Y)$; however,    more precisely,  it depends only on  $\beta$, the relative fraction, i.e., $A(Z) = A(\beta)$. 

We now proceed to prove a crucial result for the BPs, which helps in performance analysis of  
Section \ref{sec_perf_measure}. 

\section{Limit proportions of Branching process}
Transient analysis (study of growth patterns, limit proportions etc.) is an  important aspect for BPs, under super-critical regime \cite{klebaner1989geometric}. It is a common practice to scale the process appropriately that enables convergence to  a  finite  limit,  to  understand  the  otherwise transient,  exploding  process. We consider a very different type of scaling ($\Theta_n$ defined below) and   adopt  a  new  approach using \textit{stochastic approximation (SA) techniques}  (e.g. \cite{kushner2003stochastic}) to derive  (time) limit of the proportion $\beta(t)$ of the two population types   for CTMTBP with   special structure as in \eqref{eq:warning} and \eqref{Eqn_Gen_matrix}. Here,  the generator matrix $A(Z)$ depends on the population sizes only via 
$\beta$.

To this end, we analyse our process at transition epochs, let $\tau_n$ denote the  epoch at which the $n^{th}$ individual wakes up. Define $X_n:=X(\tau_n^+)$  and $Y_n := Y(\tau_n^+)$. Then, at $\tau_{n}$, if $x$-type individual wakes up (see \eqref{eqn_transition}), we obtain

\vspace{-3mm}
{\small
\begin{align*}
    X_{n} = X_{n-1} -1 + T_{F, n}^x \xi_{n}, \ \  Y_{n} = Y_{n-1} + (1- T_{F, n}^x)\xi_{n},
\end{align*}}where $\xi_{n} \stackrel{d}{=} Bin(\mathcal{F}, \eta_u)$, $T_{F, n}^x$ is defined similarly. We have similar transitions for $y$-wake up.
The total population, $S_n : = X_n + Y_n$, progresses (irrespective of the type waking up) as:

\vspace{-2mm}
{\small $$S_{n} = S_{n-1} - 1 + \xi_{n}. $$}
Let $\S_n $ represent the  sample mean formed by the i.i.d. sequence of the generated offsprings $\{\xi_n\}_n$  plus the initial populations $(x_0,y_0)$:

\Arxivtwo{
\vspace{-2mm}
{\small
\begin{align}\label{eqn_sample_mean}
    \S_n = \frac{1}{n} \left (\sum_{i=1}^n( \xi_i-1) + x_0 + y_0 \right ).
\end{align}}
}{
\vspace{-8mm}
{\small
\begin{align*}
\hspace{2.2cm}\S_n = \frac{1}{n} \left (\sum_{i=1}^n( \xi_i-1) + x_0 + y_0 \right ).
\end{align*}}}Observe that the total population ($\mu_e$ is extinction epoch),
\begin{center}
    $S_n = n\S_n 1_{n < \mu_e}$, with $\mu_e := \inf \{n : S_n = 0\}.$
\end{center}
Further, also observe that

\vspace{-2.5mm}
{\small\begin{equation}\label{Eqn_XnYnSnetc}
     X_n \le S_n \le n|\S_n| \mbox{ for all } n.  
\end{equation}}
Note that the same holds  for $Y_n$. By the strong law of large numbers,
  $\S_n \to m_\eta-1$ a.s., while $S_n/n \to m_\eta-1$ only in the survival sample paths (i.e., when $S_n > 0$  for all $n$). 
Next, we need the following notations for our analysis using SA: let $H_n$ denote the indicator that an individual of $x$-type wakes up
at the $n^{th}$ transition epoch, and $H_n^c := 1 - H_n$. Define $\Theta_n := [\psi_n, \theta_n]$ as the ordered pair respectively representing $S_n/n$ and $X_n/n$. Let $\gamma_n = 1/n, I_n := 1_{\psi_{n-1} > 0 }$.
We show that the evolution of $\Theta_n$ can be captured by the following $2$-dimensional stochastic approximation-based scheme:

\vspace{-4mm}
{\small
\begin{align}\label{eqn_SA_scheme}
        \psi_{n} &= \psi_{n-1} + \gamma_n\left(\xi_{n} - 1 - \psi_{n-1} \right)I_n \mbox{  \normalsize    and}\\
        \theta_{n} &= \theta_{n-1} + \gamma_n\left(H_n\left( T_{F, n}^x \xi_{n} - 1\right)  + H_n^c T_{F, n}^y \xi_{n}- \theta_{n-1} \right) I_n. \nonumber
        \vspace{-2mm}
\end{align}}
The analysis is derived using the SA tools of  \cite{kushner2003stochastic}. We use   similar notations  as in \cite{kushner2003stochastic}.  Define $L_n := [L_n^\psi, L_n^\theta]^T$,  where

\vspace{-4mm}
{\small
\begin{align}
    L_n^\psi &= \left(\xi_{n} - 1 - \psi_{n-1} \right)I_n \  \mbox{ \normalsize   and}\\
    L_n^\theta &= \left(H_n\left( T_{F, n}^x \xi_{n} - 1\right)  + H_n^c T_{F, n}^y \xi_{n}- \theta_{n-1} \right) I_n. \nonumber
\end{align}
}
Thus \eqref{eqn_SA_scheme} becomes  $\Theta_{n} = \Theta_{n-1} + \gamma_n L_n.$
The conditional expectation of $L_n$ with respect to $\mathcal{G}_n = \sigma \{X_k, Y_k; k \leq n\}$,   

\vspace{-4mm}
{\small
\begin{align}\label{eqn_g_bar}
 E[L_n|\mathcal{G}_n] \hspace{-4mm} &\hspace{4mm} = \bar{g}(\Theta_n), \  \mbox{\normalsize with, }  \bar{g}^\psi (\Theta) := (m_\eta-1-\psi)1_{\psi >0},\\
\hspace{-4mm}\bar{g}^\theta (\Theta)  &:= \left \{ \beta\left( q_F^u(\beta) m_\eta - 1 \right) + \left(1 - \beta\right)q_R^u(\beta) m_\eta - \theta \right \} 1_{\psi > 0 },\nonumber\\
q_i^u(\beta) &:= \alpha_i^u \left(\frac{w  \beta}{\beta + b \left(1 - \beta\right)} + \epsilon \right), \ \ \  i \in \{R, F\}. \nonumber 
\end{align}}
Now, the Ordinary Differential Equation  (ODE) 
that can  approximate \eqref{eqn_SA_scheme} is given by (see \cite{kushner2003stochastic}):
\begin{equation}\label{ODE}
    \dot{\psi} = \bar{g}^\psi (\Theta),  \mbox{\normalsize and } 
    \dot{\theta} = \bar{g}^\theta (\Theta).
\end{equation}
We prove that the ODE indeed approximates  \eqref{eqn_SA_scheme} and derive further   results mainly using \cite[Theorem 2.2, pp. 131]{kushner2003stochastic}.
Since  ${\bar g}(\cdot)$ is measurable, the results cannot be  applied directly. We provide the required justifications/modifications, identify the attractors (i.e., $(\psi^*, \theta^*)$ of Theorem \ref{thrm_ODE}) and the domain of attraction of  the ODE,  and finally derive the following result (see Appendix for proof details):

\begin{thm}\label{thrm_ODE}
Assume  {\small $E[\mathcal{F}^2] < \infty$},  $\max\{\alpha_F^u, \alpha_R^u\}(w + \epsilon) < 1$,  and $\alpha_R^u, \alpha_F^u ,  \epsilon > 0$. The sequence $(\psi_n, \theta_n)$ converges a.s. to $(\psi^*, \theta^* ) $ in co-survival sample paths, with $\psi^* = m_f \eta_u - 1$ and $\theta^* = \beta^* \psi^*$, where $\beta^*$ satisfies the fixed-point equation:
\begin{equation}\label{Eqn_theta_star}
    \beta^*  =   \beta^* q_F^u(\beta^*)  + (1 - \beta^*)q_R^u(\beta^*).
\end{equation}
Further, \eqref{Eqn_theta_star} has  unique solution in $(0,1)$. 
In other  sample paths, the sequence either converges to $(0, 0)$ (i.e., complete extinction), or $(\psi^*, \psi^*),$ or $ (\psi^*, 0)$ (i.e., only one population explodes). \eop
\end{thm}

\noindent \textbf{Remarks:} \textbf{(i)} Depending on the irreducibility of the process, the probability that the process converges to $(\psi^*, \psi^*)$ or $ (\psi^*, 0)$ could be zero; 
\textbf{(ii)} In the standard irreducible multiple type BP $(X(t), Y(t))$, it is well known that $X(t)/(X(t) + Y(t))$ converges to $1/(1 + v_y)$ a.s., where $[1, v_y]$ is the  (unique) left  eigenvector corresponding to the unique largest  eigenvalue of the generator matrix, $A$ (e.g., \cite[Theorem 2, pp. 206]{athreya}). By the direct verification, one can show that the solution of \eqref{Eqn_theta_star} and that of the following fixed point equation:

\vspace{-5mm}
{\small
\begin{eqnarray}
\hspace{1.5cm}\mbox{Left Eigenvector }  \left  ( A\left ( \frac{1}{ 1 + v_y} \right  ) \right  ) = \left[1, v_y\right], \label{Eqn_gv}
\end{eqnarray}}are connected by $\beta^* = 1/(1 + v_{ y})$. Thus, it is interesting to note that even with population dependency as in \eqref{Eqn_Gen_matrix},  the limit proportions are given by the eigenvector. However, the eigenvector is now obtained through a fixed-point equation~\eqref{Eqn_gv}.
\Arxivtwo{(iii) In the above Theorem, we implicitly assume that $m_f\eta_u > 1$, and we skipped to mention it in the published work.}{}


\Arxivtwo{ 
\subsection{Reluctant forwarding, after fake-tag}
In our warning-based mechanism, we propose users to forward the news by appropriately tagging it; however, some of the users  tend  to  be more  reluctant  to  forward  the  news  after  tagging  it as fake. One can incorporate this aspect by introducing a \textit{reluctance factor $\eta_c < 1$}; if a user tags the news as fake (real), he forwards it to $Bin(\mathcal{F}, \eta_c \eta_u)$ (respectively $Bin(\mathcal{F}, \eta_u)$) among his friends. 

We again aim to provide the limit proportions using SA techniques and the initial steps are similar to the case with $\eta_c =~1$.  We now have the following dynamics of our process for $x$-wake up at transition epoch $\tau_n$:

\vspace{-2mm}
{{\small
\begin{align*}
    X_{n} = X_{n-1} -1 + T_{F, n}^x \xi_{n, x}, \ \  Y_{n} = Y_{n-1} + (1- T_{F, n}^x)\xi_{n, y},
\end{align*}}where {\small $\xi_{n, x} \stackrel{d}{=} Bin(\mathcal{F}, \eta_u \eta_c)$, $\xi_{n, y} \stackrel{d}{=} Bin(\mathcal{F}, \eta_u)$}} and $T_{F, n}^x$ is defined similarly. Likewise the transitions for $y$-wake up can be defined. We assume $m_f \eta_u \eta_c > 1$, so that the process is in super-critical regime (the posts explode with positive probability). As defined in \eqref{eqn_sample_mean}, here we have:

\vspace{-2mm}
{\small
\begin{eqnarray}
    \S_n = \frac{1}{n}\left(\sum_{i=1}^n \left( T_{F, n}^x \xi_{n, x} - 1 +  (1- T_{F, n}^x)\xi_{n, y}\right) + x_0 + y_0\right).
\end{eqnarray}}Clearly, $\S_n$ is not the sample mean  formed by i.i.d.  offsprings, as $\{\xi_{n, x}\}_n \stackrel{d}{\neq} \{\xi_{n, y}\}_n$. However, using appropriate coupling arguments (as in the proof of Theorem \eqref{thm:monotonic_model1}), we can replace the offsprings distributed according to $\xi_{n, x}$ by those distributed as $\xi_{n, y}$.
Then the resultant sample mean, denoted by $\tilde{\S}_n$,  dominates   $\S_n$. The upper bounds of   \eqref{Eqn_XnYnSnetc} still hold and we get:
\begin{equation}\label{Eqn_XnYnSnetc_eta_c}
     X_n \le S_n \le n|\S_n|
     \le n|{\tilde \S}_n|\mbox{ for all } n.  
\end{equation}

Using the convergence of ${\tilde \S}_n$, one can proceed as in the proof of Theorem \ref{thrm_ODE};  we mention only the required changes here. The  modifications in \eqref{eqn_SA_scheme}-\eqref{eqn_g_bar} are:

{
\vspace{-4mm}
{\small
\begin{align}\label{eqn_SA_scheme_eta_c}
        \psi_{n} &= \psi_{n-1} + \gamma_n\bigg[H_n\left(T_{F,n}^x \xi_{n, x} + (1-T_{F,n}^x)\xi_{n, y} \right) - 1 - \psi_n \nonumber \\
    &\hspace{2cm}+ H_n^c\left(T_{F,n}^y \xi_{n, x} + (1-T_{F,n}^y)\xi_{n, y} \right) \bigg] I_n,\\
        \theta_{n} &= \theta_{n-1} + \gamma_n\left[H_n\left(- 1 + T_{F,n}^x \xi_{n, x} \right) + H_n^c T_{F,n}^y \xi_{n, x}  - \theta_n \right] I_n. \nonumber
        \vspace{-2mm}
\end{align}}
}Thus, \eqref{eqn_SA_scheme_eta_c} can be written compactly as $\Theta_{n} = \Theta_{n-1} + \gamma_n L_n$, where $L_n = [L_n^\psi, L_n^\theta]^T$ with:

{
\vspace{-2mm}
{\small
\begin{align}
\begin{aligned}
    L_n^\psi &= \bigg[H_n\left(T_{F,n}^x \xi_{n, x} + (1-T_{F,n}^x)\xi_{n, y} \right) - 1 - \psi_n  \\
    &\hspace{1.2cm}+ H_n^c\left(T_{F,n}^y \xi_{n, x} + (1-T_{F,n}^y)\xi_{n, y} \right) \bigg] I_n, \\
    L_n^\theta &= \left[H_n\left(- 1 + T_{F,n}^x \xi_{n, x} \right) + H_n^c T_{F,n}^y \xi_{n, x}  - \theta_n \right] I_n. 
    \end{aligned}
\end{align}
}
}
We have the ODE as in \eqref{ODE} with $\bar{g}^\psi(\Theta), \bar{g}^\theta(\Theta)$ as:

\vspace{-4mm}
{\small
\begin{align}\label{eqn_g_bar_eta_c}
\begin{aligned}
\bar{g}^\psi (\Theta) &:= \bigg(m_\eta (\eta_c - 1) q_F^u(\beta)\beta + m_\eta - 1 - \psi \\
&\hspace{1.2cm}+ m_\eta (\eta_c - 1) q_R^u(\beta)\left(1 - \beta\right) \bigg)1_{\{\psi > 0 \}}, \\
\bar{g}^\theta (\Theta)  &:= \bigg ( \beta\left(- 1 + q_F^u(\beta) m_\eta \eta_c \right) \\
&\hspace{1.2cm}+\left( 1 - \beta\right)q_R^u(\beta) m_\eta \eta_c  - \theta \bigg )1_{\{\psi > 0 \}} .
\end{aligned}
\end{align}}
The rest of the SA-based details are the same, but the analysis of the ODE \eqref{eqn_g_bar_eta_c} is different.  We provide the ODEs based analysis to complete the proof of the following result in Appendix:

\begin{thm}\label{thrm_ODE_eta_c}
Assume  $E[\mathcal{F}^2] < \infty$,  $\max\{\alpha_F^u, \alpha_R^u\}(w + \epsilon) < 1$
and $\alpha_R^u, \alpha_F^u ,  \epsilon > 0$. The sequence $(\psi_n, \theta_n)$ converges a.s. to $(\psi^*, \theta^* ) $ in co-survival sample paths, with $\theta^* = \beta^* \psi^*$, where  
$$
\psi^* = m_\eta - 1 - m_\eta (1 - \eta_c) \left[q_F^u(\beta^*)\beta^* +  q_R^u(\beta^*)\left(1 - \beta^*\right)\right] 
$$
and $\beta^*$ satisfies the fixed-point equation:

\vspace{-2mm}
{\small 
\begin{equation}\label{Eqn_theta_star_eta_c}
    \left[\beta^* q_F^u(\beta^*) + \left(1-\beta^* \right) q_R^u(\beta^*)\right] \left[\eta_c + \beta^*(1 - \eta_c) \right] = \beta^*.
\end{equation}}Further, \eqref{Eqn_theta_star_eta_c} has a unique solution in $(0,1)$. 
In the other   sample paths, the sequence either converges to $(0, 0)$ (i.e., complete extinction), or $(\psi^*, \psi^*),$ or $ (\psi^*, 0)$ (i.e., only one population explodes). \eop
\end{thm}

\noindent \textbf{Remarks:} Even with the reluctance factor, the limit proportion converges to a  unique limit, the  unique zero of \eqref{Eqn_theta_star_eta_c}. The sum population also converges to a unique limit, $\psi^*$, which now depends on $\beta^*$ (unlike the one in Theorem \ref{thrm_ODE}).

The rest of the paper is dedicated to understand the case with $\eta_c = 1$ and we plan to study other aspects in the presence of reluctance factor in future.
}
{ }

\section{Performance Measures}\label{sec_perf_measure}
We aim to control the warnings $\omega_t$  (through  parameters $w$,  $b$) to mitigate fake news propagation, without significantly affecting the propagation of real news. To this end, we appropriately define type-$1$ and type-$2$ performances. \textit{The type-$1$ performance quantifies the effects of controlled warnings on the fake-news propagation, whereas the type-$2$ performance quantifies the adverse effects on the real-news propagation}.

It is  important to ensure that most users are informed of the fake news, when $u = F$. Thus, we define the type-$1$ performance ($\Psi_1$) as the (time) asymptotic fraction of posts with real-tag $Y(t)$:

\vspace{-5mm}
{\small\begin{equation} \label{eq:typ1_TaF_1}
    \hspace{1.9cm}\Psi_1(w, b) :=  \lim_{t \to \infty}  \left( \frac{Y(t)}{X(t)+Y(t)}\right)_{u=F}.
\end{equation}}
By  Theorem \ref{thrm_ODE}, 
$ \Psi_1 = \left(1 - \beta^* \right)_{u=F}$.  
When the underlying news is fake, minimizing $\Psi_1$ ensures that a minimum number of users are mis-informed about the news being real. 

In a similar way, when the underling news is real, we define the type-$2$ performance ($\Psi_2$) as the asymptotic fraction of posts with fake-tag, which is also provided by Theorem~\ref{thrm_ODE}:

\vspace{-4mm}
{\small\begin{equation} \label{eq:type2_TaF_1}
\Psi_2(w,b) := \lim_{t\to \infty}\left( \frac{X(t)}{X(t)+Y(t)}\right)_{u=R} = \left( \beta^* \right)_{u=R}.
\end{equation}}
Ensuring $\Psi_2$ is within a given limit gives an  upper bound on the number of real news copies mis-tagged as fake.

Both $\Psi_1$ and $\Psi_2$  are non-negative fractions  bounded by $1$. We immediately have following properties with respect to the control parameters, $(w, b)$ (See proof in the Appendix):  
\begin{thm} \label{thm:monotonic_model1}
Assume $\alpha^u_F > \alpha^u_R$. 
When the  systems start with  same initial state,
	the type-$1$ performance ($\Psi_1$) decreases  and the type-$2$ performance ($\Psi_2$) increases, monotonically with  increase in $w$. The same is true for a decrease in $b$.
	\hfill\eop	
\end{thm} 
\noindent This theorem is useful for the optimal design of  next section.

\section{Optimal warning  parameters}\label{section_5}
In practice, adverse effects on real news propagation cannot be  allowed beyond a certain tolerance threshold. We thus impose a constraint that   $\Psi_2$ is upper bounded by $c$, where $c >0$ is a design parameter.
  We consider the   following optimization problem\footnote{
One can have a sense of the comparison of various parameters like $\{\alpha_i^u\}_{i,u}$ (as explained in   Section \ref{sec_regime}), but may not have an exact estimate of the parameters. We assume, any $w \le 1$ ensures probability $q_i^u (\beta) \le 1$ for all possible $i, u, \beta$ and parameters. 
}   for the optimal design:
\begin{eqnarray}
 \min_{w,b}  \Psi_1(w,b)  \hspace{2mm} 
\mbox{s.t.  } \Psi_2(w,b) \leq c; \ 
 0 \leq w \leq 1; \ b \geq 0. \label{opt_problem}
\end{eqnarray}
 By Theorem \ref{thm:monotonic_model1}, $\Psi_2(0,1)< \Psi_2 (1, 0)$. Further \eqref{opt_problem} simplifies to (Proof in Appendix):
\begin{lem}\label{lem:opt}
Assume that {$\Psi_2(0,1)\hspace{-0.5mm}<\hspace{-0.5mm}c\hspace{-0.5mm}<\hspace{-0.5mm}\Psi_2 (1, 0)$}, which implies non-empty feasible region. Then, the
optimal value of optimization problem  (\ref{opt_problem}) is achieved when $\Psi_2=c$. \eop
\end{lem}

\noindent Thus the feasible region reduces to $\{(w, b) :  \Psi_2 (w,b)= c\}$. Hence, by virtue of Lemma \ref{lem:opt} and using \eqref{Eqn_theta_star}, we can express the variable $b$ as a function of $w$  (${\bar \alpha}_c^R := (c\alpha^{R}_F+(1-c)\alpha^{R}_R)$):

\vspace{-4mm}
{\small
\begin{eqnarray}\label{opt_b_w}
b(w) =  \frac{c}{1-c}  \frac{(w+\epsilon) {\bar \alpha}_c^R-c}{c-\epsilon {\bar \alpha}_c^R }  \mbox{,  }   \ 
 \frac{\partial b(w)}{\partial w} =   \frac{c}{1-c}  \frac{{\bar \alpha}_c^R}{c-\epsilon {\bar \alpha}_c^R}.
 \end{eqnarray}} From above, there exists at maximum   one  $b(w)$ for any $ w$,  that together satisfy $\Psi_2 = c$. Thus, $w \mapsto b(w)$ is a well-defined function on $\{w: b(w) \geq 0 \}$ and 
thus the optimization problem  \eqref{opt_problem} reduces to $1$-dimensional problem:
 \begin{eqnarray}
 \min_{w}  \Psi_1(w,b(w))  \hspace{2mm} 
\mbox{s.t.  } 
 0 \leq w \leq 1; \ b(w) \geq 0. \label{opt_problem_1D}
 \end{eqnarray}
By differentiating both sides of \eqref{Eqn_theta_star}  by $w$,   the  derivative, $ {\partial  \psi_1}/{\partial w} = {\partial  \beta^*}/{\partial w}$, satisfies  the fixed-point equation:

\vspace{-4mm}
{\small
\begin{align*}\frac{\partial \beta^*}{\partial w} &= (1-\beta^*) \frac{\partial q_R^u(\beta^*)}{\partial w} + \beta^* \frac{\partial q_F^u(\beta^*)}{\partial w} \\
&\hspace{-8mm}+  \frac{\partial \beta^*}{\partial w} \left( q_F^u(\beta^*) + \beta^* \frac{\partial q_F^u(\beta^*)}{\partial \beta^*} -  q_R^u (\beta^*)  + (1-\beta^*) \frac{\partial q_R^u(\beta^*)}{\partial \beta^*}\right).
\end{align*}}
Similarly, one can derive $  {\partial \beta^*}/{\partial b}$.
Observe that the partial derivative (see \eqref{opt_b_w}):
\vspace{-4mm}
$$
\hspace{27mm} 
\frac{\partial \Psi_1}{\partial w} =  \left (
-\frac{\partial \beta^*}{\partial w} - \frac{\partial \beta^*}{\partial b} \frac{\partial b}{\partial w} \right )_{u = F}.
$$
Using these derivatives, we can solve \eqref{opt_problem_1D}  by the following projected gradient descent algorithm:
$$w_{l+1} = \left[ \left | w_l - \kappa_l \frac{\partial \Psi_1}{\partial w} \right |_{(w, b) = (w_l, b(w_l) ) } \right]_{[0,1]\cap \{b(w) \geq 0\}}, $$
where $[  \cdot ]_{\mathcal{A}}$ is the projection to set $\mathcal{A}$, and $\{\kappa_l\}$ is a decreasing sequence of step sizes. We obtain the optimizers, $(w^*, b(w^*))$, and the optimal $\Psi_1$ for several numerical examples using the above method in the following section. 

Before proceeding further, we discuss some   meaningful assumptions,    naturally required for the application. 

\subsection {Suitable Regime for Parameters:}
\label{sec_regime}
Users may sometimes find  fake news more attractive (\cite{Vosoughi1146});  
 therefore, we assume that $\eta_F > \eta_R$.
When the underlying news $u$ is the same, we assume that $\alpha_F^u > \alpha_R^u$, which indicates that the probability of a user tagging the news as fake is higher when the sender's tag is fake. \textit{We model the intrinsic capability of users to recognize the veracity of the news by assuming $\alpha_F^F > \alpha_F^R$}. This assumption indicates that users are more likely to tag fake   news as fake, as compared to tagging real   news as fake.

\section{Numerical Observations}
We corroborate the results of Theorem \ref{thrm_ODE} using exhaustive Monte-Carlo (MC) simulations for different sets of parameters. In Table \ref{tab:conjecture}, one can observe that the relative proportions (i.e., $X/(X+Y)$) well match with $\beta^*$ of Theorem \ref{thrm_ODE}.

\begin{table}[htbp]
\vspace{1mm}
\centering
\begin{tabular}{|c|c|c|c|c|} 
\hline
                                                               &                      & Config 1          & Config 2          & Config 3           \\ 
\hline
\multirow{2}{*}{Parameters}                                    & $\alpha^u_F,\ \eta_u$  & 0.9, 0.3          & 0.5, 0.15         & 0.85, .35          \\ 
\cline{2-5}
                                                               & $\epsilon,\ b$       & 0.05, 1           & 0.05, 1           & 0.1, 0.5           \\ 
\hline
\begin{tabular}[c]{@{}l@{}}Simulations\end{tabular}  & ${X}/{(X+Y)}$           & \textbf{0.04434 }  & \textbf{0.0171}  & \textbf{0.39464 }   \\ 
\hline
\begin{tabular}[c]{@{}l@{}}Solution of \eqref{Eqn_theta_star}\end{tabular} & $\beta^*$    & \textbf{0.04425 }  & \textbf{0.01701}   & \textbf{0.39533 }   \\
\hline
\end{tabular}
\caption{Sample path   results  with $m_f=~30, \ w=1,   \lambda=0.1,\ \alpha_R^u = 0.5\times\alpha_F^u$.}
\label{tab:conjecture}
\vspace{-3mm}
\end{table}
\vspace{1mm}
\noindent\textbf{Validation}:
We validate theoretical results for $\Psi_1$ and $\Psi_2$  by MC simulations based on ego-network dataset of Twitter provided by \textit{SNAP} (\cite{snapnets}). It consists of $81,306$ users and $1,768,149$ (directed) connections among them.

\begin{figure}[htbp]
 \vspace{1mm}
     \centering
     \begin{subfigure}[b]{0.22\textwidth}
         \centering         \includegraphics[width=1.0\textwidth]{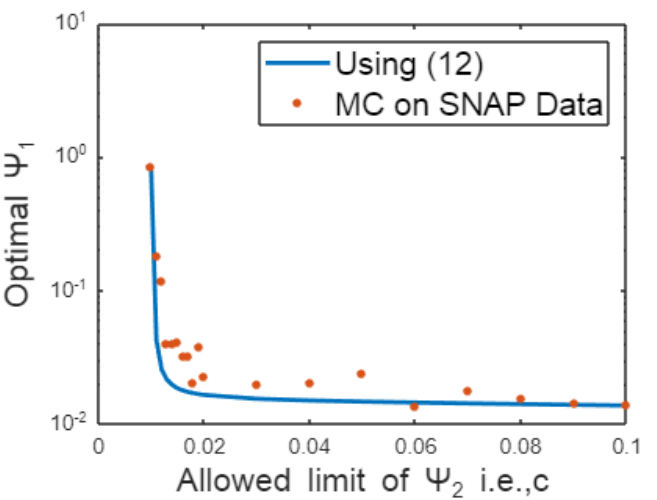}
         \caption{Validation, real data }
    \label{fig:snap_TaF}
     \end{subfigure}
     \hspace{4mm}
     \begin{subfigure}[b]{0.22\textwidth}
         \centering
        \includegraphics[width=1.0\textwidth]{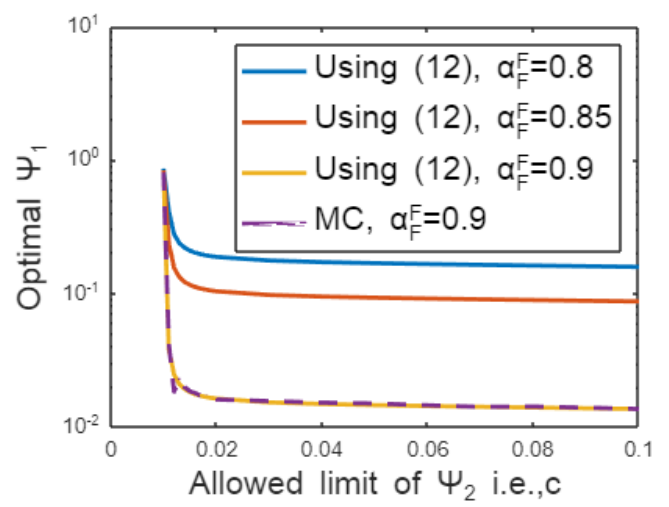}
         \caption{Optimal $\Psi_1$} 
    \label{fig:opt_type1_spread}
     \end{subfigure}
     \hfill
      \caption  {$\mbox{Common Parameters: } \eta_F=.08,\ \eta_R=.05,\ m_{f}=28,\ \lambda=.1,\ \alpha_R^F=.75\times\alpha_F^F,\ \alpha_F^R=.3,\ \alpha_R^R=.09,\  \epsilon=.1$.}
        \label{fig:three graphs}
        \vspace{-4mm}
\end{figure}

Each run of the simulation begins with  two users, one with fake tag and one with real tag, initial users are chosen randomly from SNAP data-set. We generate exponential random variable (RV) (with parameter $\lambda(X_n + Y_n)$) to represent the inter-visit time of the $n^{th}$ user to OSN.  At each user's visit,  user tags the news  (as explained in section \ref{Sec_System_model}) using binary RV,   while  the news is shared   to  any  of its connections (as given by SNAP data-set),       independently of others with probability $\eta_u$. 
 We use   optimal warning parameters and use average number of connections of data as $m_f$ for  theoretical expressions. We generated 20 such sample-paths/runs, each of which stop after system time\footnote{Around 3500-5000 unread copies of post are shared before    time $t=30$. } (updated by the inter-visit times of the users) equal to 30.  
 The MC performances obtained  by  averaging over 20 such sample-paths  closely match with asymptotic proportions obtained using \eqref{Eqn_theta_star} (see Figure~\ref{fig:snap_TaF}).

\vspace{-4mm}
\vspace{5mm}
\noindent\textbf{Performance  at Optimal warning}: 
In Figure~\ref{fig:opt_type1_spread}, we plot optimal type-$1$ performance as a function of  threshold/tolerance value of type-$2$ performance (see \eqref{opt_problem}). In the same figure, we also plot results for type-$1$ performance obtained from MC simulations. One can observe that  the limiting theoretical results match exactly with chosen MC simulations sample-path-wise. 

From Figure~\ref{fig:opt_type1_spread}, the control mechanism significantly improves the system performance for all values of $c$.
For an instance, if $2\%$ tolerance  is chosen for the  type-$2$ performance (i.e., $c = 0.02$ in \eqref{opt_problem}),  
only $10\%$ of the OSN users are mis-informed   (and mis-tag) about the fake news being real, when $\alpha_F^F = 0.85$. In contrast,  
for the same settings with no controlled warnings,  $72\%$ users mis-tag the fake news as real; here, the users tag the news based on constant warning level, $\epsilon$ (see \eqref{eq:warning}) and their intrinsic capacity.

\section{Conclusions and Future Work}
We have considered the problem of mitigating fake news in online social networks without affecting the propagation of authentic ones.  We have designed a warning-based control mechanism, where the warning depends on the responses of forwarding users. 
The spread of fake news is modeled using an appropriate population size-dependent multi-type continuous-time branching process.

Our contributions are two-fold. On one hand, we have results towards the above mentioned special branching processes. On the other hand, we have proposed a  warning-based control mechanism. 
We have adopted a   new type of scaling and used stochastic approximation  tools to derive
 time-asymptotic proportions of the individual populations of the size-dependent branching process. These proportions are represented as the solution of a simple fixed-point equation that depends upon system parameters.
Using the new  results in branching process, we computed different performance measures of the spreading (under warning) to capture the effectiveness of the control mechanism. We  identified structural (monotone) properties of the performance measures. 

Finally, we have formulated an optimization problem, where the network controller optimizes the type-$1$ (fake news) performance subject to a constraint on the degradation of the type-$2$  (authentic news) performance. With $1\%, 1.5\%$ and $2\%$   degradation on authentic news, $16\%, 88\%$ and $90\%$  users, respectively, identify the fake news as fake. 
 
\Arxivtwo{}{
\noindent \textbf{Reluctant forwarding, after fake-tag:}
In our warning-based mechanism, we propose users to forward the news by appropriately tagging it; however, some of the users  tend  to  be more  reluctant  to  forward  the  news  after  tagging  it as fake. One can incorporate this aspect by introducing a \textit{reluctance factor $\eta_c < 1$}; if a user tags the news as fake (real), he forwards it to $Bin(\mathcal{F}, \eta_c \eta_u)$ (respectively $Bin(\mathcal{F}, \eta_u)$) among his friends. 

We  have some initial analysis in \cite{arxiv}; we have a result exactly as in  Theorem \ref{thrm_ODE}, which differs only in the co-survival limits:

\vspace{-4mm}
{\small $$\psi^* = m_\eta (\eta_c - 1) \left[ q_F^u(\beta^*)\beta^* +  q_R^u(\beta^*)\left(1 - \beta^*\right) \right] + m_\eta - 1,$$}
and $\beta^*$ now satisfies the fixed-point equation:

\vspace{-4mm}
{\small\begin{equation}\label{Eqn_theta_star_eta_c}
    \left[\beta^* q_F^u(\beta^*) + \left(1-\beta^* \right) q_R^u(\beta^*)\right] \left[\eta_c + \beta^*(1 - \eta_c) \right] = \beta^*.
\end{equation}}
We aim to investigate the influence of $\eta_c$ on our warning-based mechanism in  future.
}


\section*{Appendix}


\noindent \textbf{ Proof of Theorem \ref{thrm_ODE}:}
 We  prove the result using   \cite[Theorem 2.2, pp. 131]{kushner2003stochastic}, as  ${\bar g}(\cdot)$ is only measurable. Towards this, we first need to prove (a.s.) equicontinuity of sequence $\Theta^n(t) := \Theta_n + \sum_{i=n}^{m(t_n+t)-1}\epsilon_i L_i$, with $m(t) = \max\left  \{ n: \sum_{k=0}^{n-1}  \gamma_k \le t \right \}$. This proof goes through exactly as in the proof of   \cite[Theorem 2.1, pp. 127]{kushner2003stochastic} because of the following reasons:  the random vector  $L_n$ is comprised of $\theta_n, \psi_n$ and i.i.d. random variables and by \eqref{Eqn_XnYnSnetc}, it suffices to show that $\sup_n E|\S_n|^2 < \infty$, which is trivially true because $E[\mathcal{F}^2] < \infty$;
further, we exactly have $E[L_n|{\cal F}_n] = {\bar g} (\Theta_n)$ (here $\beta_n$ in
\cite[Assumption {\bf A}.2.2]{kushner2003stochastic} is 0), as well the projection term $Z_n \equiv  0$. Further, $\{\Theta_n(0)\}_n$ is bounded a.s. by strong law of large numbers as applied to $\{\S_n\}_n$.

In  Lemma \ref{lemma_ASL}, we identify  the  attractors\footnote{ A set $A$ is said to be Asymptotically stable in the sense of Liapunov, if there exist a neighbourhood (called domain of attraction, D($A$)) starting in which the ODE trajectory converges to $A$ as time progresses (e.g., \cite{kushner2003stochastic}). } of \eqref{ODE}, with $\theta^*$  as in \eqref{Eqn_theta_star}. 
Proof is now completed  sample-path wise.   

First consider the sample-paths in which $\psi_n \to 0$ (i.e., $S_n \to 0$). Then clearly, $(\psi_n, \theta_n) \to (0, 0)$. For the sample paths such that $(\psi_n, \theta_n)$ converges to $(\psi^*, \psi^*)$ or $(\psi^*, 0)$, there is nothing left to prove. In the remaining sample paths, $\psi_n \to \psi^*$ a.s. (where $\psi^* = m_\eta - 1$ as in \eqref{Eqn_theta_star}). Further, $(\psi_n, \theta_n)$ visits $S_\delta$ of Lemma \ref{lemma_ASL} (for any $0 < \delta < \psi^*$) infinitely often.  By applying \cite[Theorem 2.2, pp. 131]{kushner2003stochastic}  to these sample paths, the sequence   converges to $(\psi^*, \theta^*)$.  \eop 

\begin{lem}\label{lemma_ASL} 
For ODE \eqref{ODE}, $(\psi^*, \theta^*)$ is asymptotically stable in the sense of Liapunov. For any $0 < \delta < \psi^*$, the set\footnote{Define $\overline{\N_\delta} (\psi^*) := \{\psi: |\psi-\psi^* | \leq \delta\}$.}

\vspace{-2mm}
{\small
$$
S_\delta  = \left\{(\psi, \theta): \psi \in \overline{\N_\delta}(\psi^*), \frac{\theta}{\psi} \in [\delta,1- \delta] \right\} ,
$$}is  compact 
and is in the domain of attraction of $(\psi^*, \theta^*)$.
\end{lem}
\noindent \textbf{Proof:}
The $\psi$-component of \eqref{ODE} has the following solution:
 
 \vspace{-4mm}
 {\small 
 \begin{eqnarray}\label{eqn_psi} 
 \psi(t) =
\begin{cases} 
 e^{-t}(\psi(0) - m_\eta + 1)+ m_\eta -1,\hspace{-2mm}  & \mbox{\normalsize if } \psi(0) > 0, \\
\psi(0), &\mbox{\normalsize if }\psi(0) \leq 0 .
\end{cases}
\end{eqnarray}}
Thus, $\psi^*  = m_\eta = m_f \eta_u-1$ is asymptotically stable with $(0, \infty)$ as domain of attraction. For $\theta$ component,  one needs to substitute solution $\psi(t)$ in its ODE (${\bar g}^\theta$ of \eqref{ODE}) to analyze. Clearly, $\theta^* = \psi^* \beta^*$, with $\beta^*$  a solution to  \eqref{Eqn_theta_star}, is an equilibrium point\footnote{In this context, the point $\bar{\theta}$ is an equilibrium point if ${\bar g}^\theta (\psi^*, {\bar \theta}) = 0$. }.

We prove the stability of the above equilibrium point using the ODE corresponding to $\beta = \theta/\psi$ (using \eqref{ODE}):

\vspace{-3mm}
{\small 
\begin{align}\label{eqn_beta_dot}
\hspace{-3mm}\dot{\beta} = \frac{\dot{\theta}}{\psi} - \frac{\theta}{\psi^2}\dot{\psi} =  1_{\psi > 0 }\frac{m_f \eta_u}{\psi} g_\beta(\beta),
\end{align}}where, $g_\beta(\beta) := \beta\left(q_F^u(\beta) - q_R^u(\beta) - 1 \right) + q_R^u(\beta)$. Note that, $g_\beta(\cdot)$ is strict convex function of $\beta$ when $b > \frac{\alpha_R}{\alpha_F},$ strict concave when $b < \frac{\alpha_R}{\alpha_F}$ and linear otherwise, because, the second derivative,

\vspace{-9mm}
\begin{align*}\hspace{24mm}
 \hspace{2mm} \ddot{g}_\beta(\beta) = \frac{2wb}{(\beta + b(1-\beta))^3} (b \alpha_F^u - \alpha_R^u).
\end{align*}

\noindent Secondly,
{\small $g_\beta(0)= \alpha_R^u \epsilon > 0$}, {\small$g_\beta(1) = \alpha_F^u(w + \epsilon) - 1 < 0$}. Using the above two, we have: (i) $g_\beta$ has unique zero, $\beta^*$, which satisfies \eqref{Eqn_theta_star}; and (ii) $t \mapsto \beta(t)$ is strictly increasing  (derivative, $g_\beta(\beta)$, strictly positive) when $0 < \beta(t) < \beta^*$, and strictly decreasing when  $1 > \beta(t) > \beta^*$. Thus from \eqref{eqn_beta_dot}, for any initial condition $(\psi_0, \theta_0) \in S_\delta$, $\beta (t) \stackrel{t \to \infty}{\to} \beta^*$ ($\psi$  given by \eqref{eqn_psi}).  \eop

\Arxivtwo{
\vspace{0.2cm}
\noindent \textbf{ Proof of Theorem \ref{thrm_ODE_eta_c}:} 
The proof of Theorem \ref{thrm_ODE_eta_c} can be done using \cite[Theorem 1(ii)]{agarwal2021new} - observe that the assumptions \textbf{A.1}-\textbf{A.2} (in \cite{agarwal2021new}) are trivially true (see \eqref{eqn_offspring_x}, \eqref{eq:transition2}), and \textbf{A.3}-\textbf{A.4} are true by Lemma \ref{lemma_ASL2}.  \eop 

\begin{lem}\label{lemma_ASL2}
For ODE \eqref{ODE}, $(\psi^*, \theta^*)$ given in Theorem \ref{thrm_ODE_eta_c} is asymptotically stable in the sense of Liapunov. For any $0 < \delta < \psi^*$, the set\footnote{Define $\overline{\N_\delta} (\psi^*) := \{\psi: |\psi-\psi^* | \leq \delta\}$.}

\vspace{-2mm}
{\small
$$
S_\delta  = \left\{(\psi, \theta): \psi \in \overline{\N_\delta}(\psi^*), \frac{\theta}{\psi} \in [\delta,1- \delta] \right\} ,
$$}is  compact 
and is in the domain of attraction of $(\psi^*, \theta^*)$.
\end{lem}
\noindent \textbf{Proof:} Similar to Lemma \ref{lemma_ASL}, consider the ODE corresponding to $\beta = \theta/\psi$ (using \eqref{eqn_g_bar_eta_c}, \eqref{ODE}):
\vspace{-1mm}
{\small 
\begin{align}\label{eqn_beta_dot_eta_c}
\hspace{-3mm}\dot{\beta} &= \frac{\dot{\theta}}{\psi} - \frac{\theta}{\psi^2}\dot{\psi} =  1_{\psi > 0 }\frac{m_\eta}{\psi} g_\beta(\beta), \mbox{ where}\\
g_\beta(\beta) &:= \left( \beta\left(q_F^u(\beta) - q_R^u(\beta) \right) + q_R^u(\beta)\right) \left(\eta_c - \beta(\eta_c -1) \right) - \beta \nonumber
\end{align}}
For $\beta^*$ a solution to  \eqref{Eqn_theta_star_eta_c} (i.e., zero of $g_\beta(\beta)$), we have $\psi^*$ (as in Theorem \ref{thrm_ODE_eta_c}) and $\theta^* = \psi^* \beta^*$ is an equilibrium point. 

Note that, $g_\beta(\cdot)$ is strict convex function of $\beta$ when $b\alpha_F^u \geq \alpha_R^u$, because, second derivative in this case:

\vspace{-2mm}
{\small
\begin{align*}
    \ddot{g}_\beta(\beta) &= \frac{d^2 g_\beta(\beta)}{d^2 \beta} = 2(1 - \eta_c) \left(q_F^u(\beta) - q_R^u(\beta) \right)\\
    &\hspace{10mm}+\frac{2wb}{(\beta + (1-\beta)b)^3} \bigg[\left(b\alpha_F^u - \alpha_R^u  \right)\left(\eta_c(1 - \beta) +\beta \right)\\
    &\hspace{10mm}+ (1-\eta_c ) \left(\beta(\alpha_F^u-\alpha^u_R) + \alpha^u_R \right)(\beta + (1-\beta)b)   \bigg] ,
\end{align*}}is positive. The rest of the proof for $b \alpha_F^u \geq \alpha_R^u$ follows as in Lemma \ref{lemma_ASL}.

\begin{figure*}[h]
\begin{minipage}{.3\textwidth}
 \centering
    \def\svgwidth{\columnwidth}
    \resizebox{0.8\textwidth}{!}{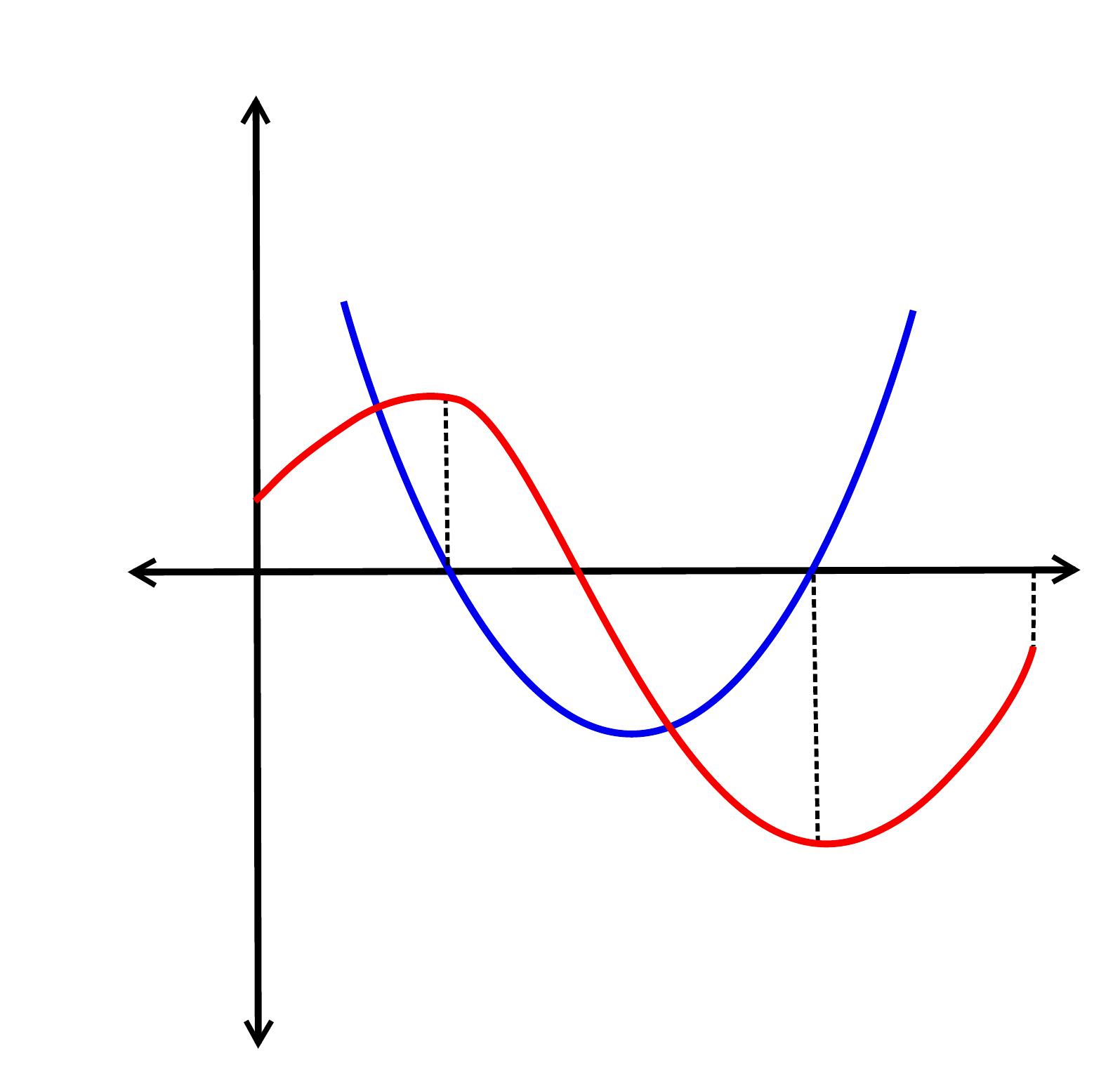}
\end{minipage}%
\hspace{10mm}
\begin{minipage}{.3\textwidth}
 \centering
    \def\svgwidth{\columnwidth}
    \resizebox{0.8\textwidth}{!}{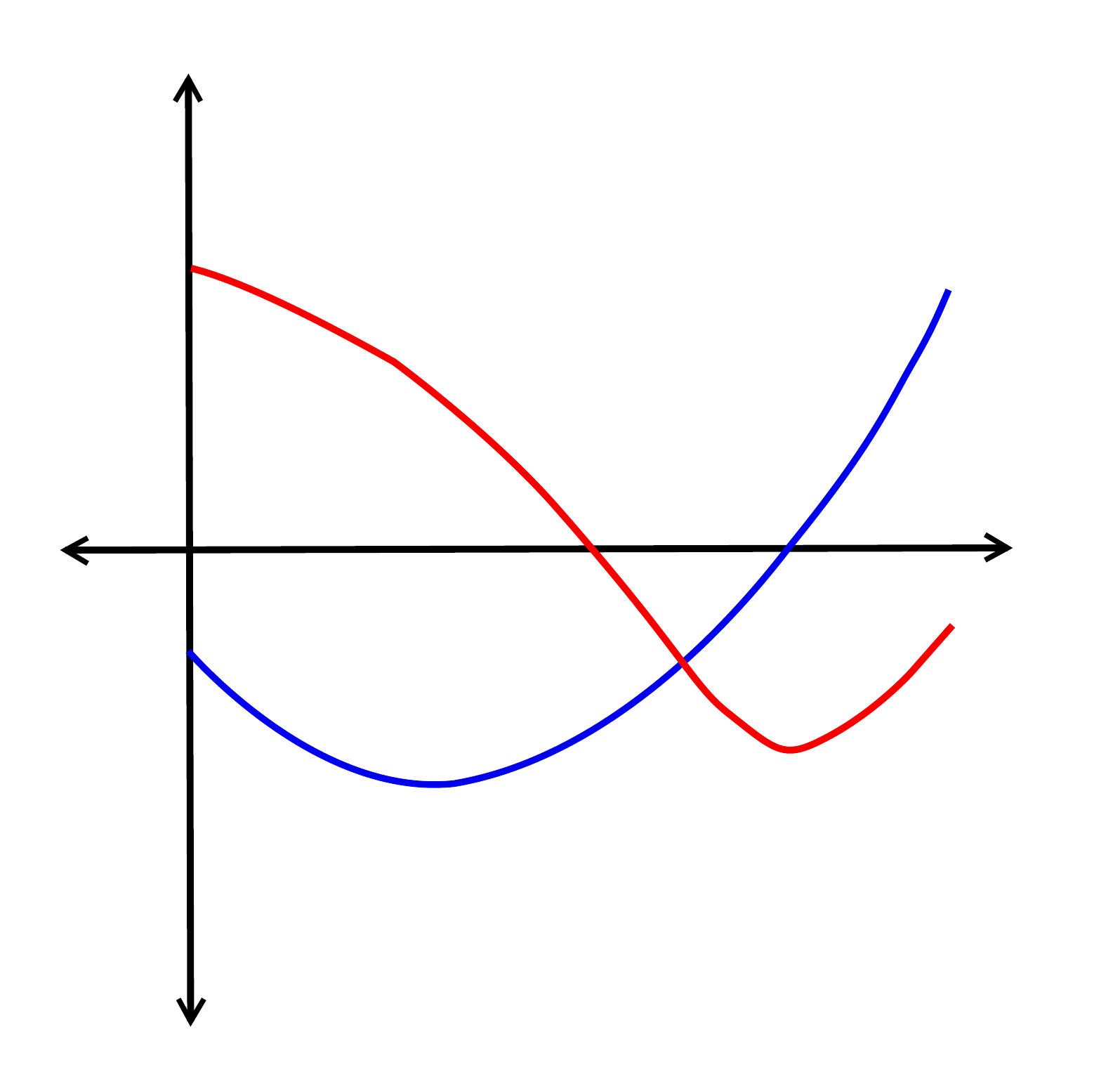}
\end{minipage}%
\hspace{10mm}
\begin{minipage}{.3\textwidth}
 \centering
    \def\svgwidth{\columnwidth}
   \resizebox{0.8\textwidth}{!}{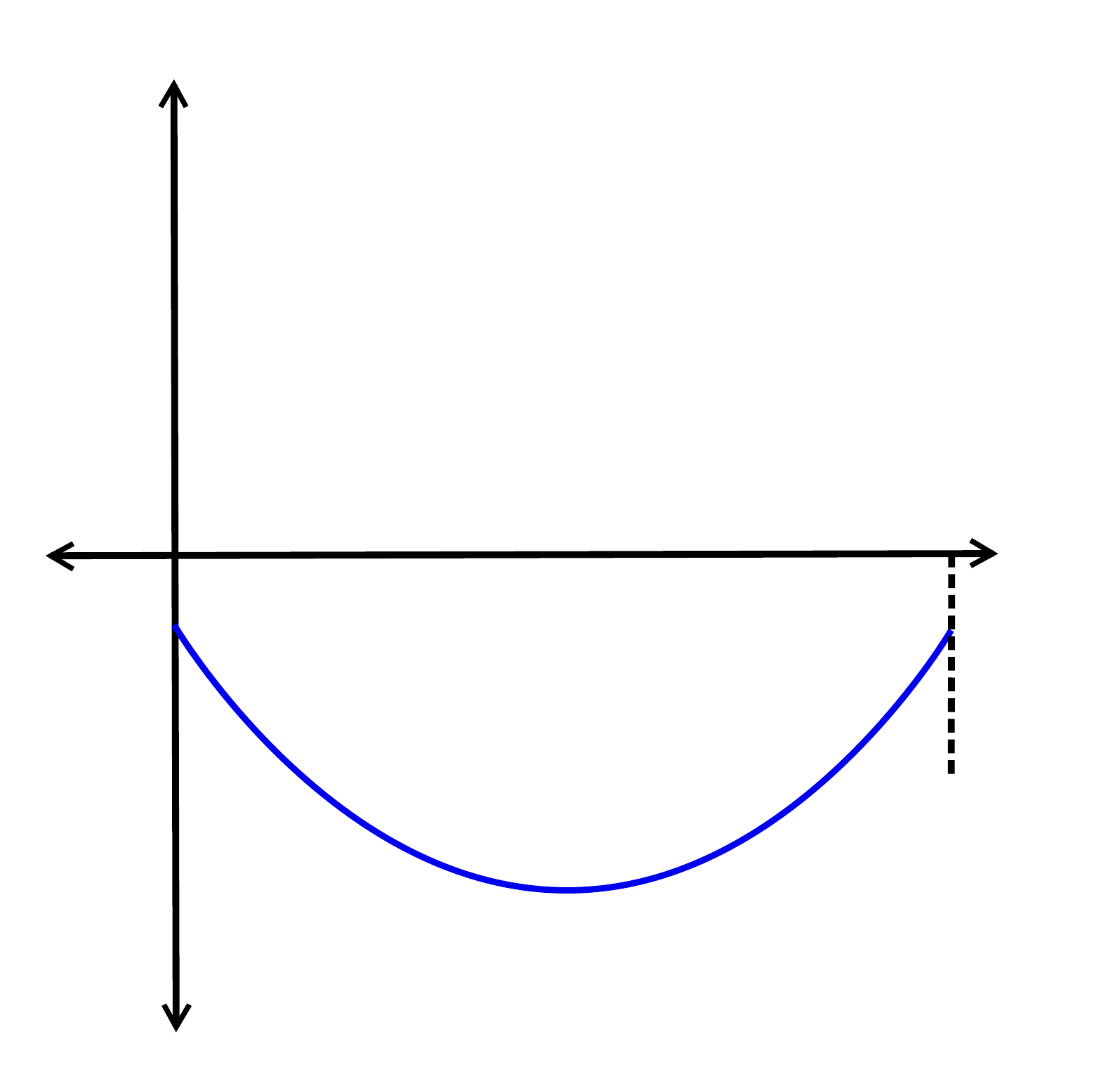}
\end{minipage}
\caption{The functions $g_\beta(\cdot)$ and $\dot{g}_\beta(\cdot)$, when $ b \alpha_F^u < \alpha_R^u \mbox{ and } b < \eta_c$
\label{figure_proof} }
\end{figure*}

For $b \alpha_F^u < \alpha_R^u$, consider the third derivative of $g_\beta(\beta)$:

\vspace{-2mm}
{\small\begin{align*}
    \dddot{g}_\beta(\beta) &= \frac{2wb}{(\beta + (1-\beta)b)^4}k(\beta), \mbox{ where }\\
    k(\beta) &= (b\alpha_F^u-\alpha_R^u)[(b-1) (2\beta(1-\eta_c) + 3\eta_c) + b(1-\eta_c)]  \\
    &\hspace{0.4cm}+  2(1-\eta_c)(\beta(\alpha_F^u-\alpha_R^u) + \alpha_R^u)(b-1)(\beta + (1-\beta)b)  \\
    &\hspace{0.4cm}+ 2(1-\eta_c)(\alpha_F^u-\alpha_R^u)(\beta + (1-\beta)b)^2.
\end{align*}}Now, $\dddot{g}_\beta(\beta)$ is a product of two terms, which has positive first term. Further, it is easy to verify $\frac{d k(\beta)}{d \beta} = 0$, which implies that $k(\beta)$ is a constant for all values of $\beta$. Notice that $k(\beta) = k(1) = 3(b\alpha_F^u-\alpha_R^u)(b-\eta_c)$ for all $\beta \in [0, 1)$.
Now, we will analyse the following cases:

$\bullet$ For $b = \eta_c$: we have $\dddot{g}_{\beta}(\beta) = 0$, which implies that $\ddot{g}_{\beta}(\cdot)$ is a constant function of $\beta$. We have, for all $\beta \in [0,1)$:
\begin{align*}
    \ddot{g}_{\beta}(\beta) = \ddot{g}_{\beta}(1) = 2(\alpha_F^u-\alpha_R^u)(wb + \epsilon(1-\eta_c)) > 0,
\end{align*}
which implies $g_\beta(\cdot)$ is a convex function of $\beta$.

$\bullet$ For $b > \eta_c$: we have $\dddot{g}_{\beta}(\beta) < 0$, which implies $\ddot{g}_{\beta}(\beta)$ is a decreasing function of $\beta$, and  for all $ \beta \in [0, 1)$:

\vspace{-2mm}
{\small
$$\ddot{g}_{\beta}(\beta) \geq \ddot{g}_{\beta}(1) = 2(\alpha_F^u-\alpha_R^u)(wb + \epsilon(1-\eta_c)) + 2wb \alpha_F^u (b-\eta_c) > 0. $$}Thus, $\ddot{g}_{\beta}(\beta) > 0$ for all $\beta$. This implies that $g_\beta(\beta)$ is a convex function. 

Hereafter, the proof  follows as in Lemma \ref{lemma_ASL} for the above discussed cases, i.e., $b \geq \eta_c$.

$\bullet$ For $b < \eta_c$:
we have $\dddot{g}_{\beta}(\beta) > 0$, which implies that $\dot{g}_{\beta}(\beta)$ is convex function of $\beta$. Now, notice that $\dot{g}_\beta(\cdot)$ can not be non-negative for all values of $\beta$, as it would imply $g_\beta(\cdot)$ is a non-decreasing  function of $\beta$, which contradicts the fact that {\small $g_\beta(0)= \eta_c \alpha_R^u \epsilon > 0$}, {\small$g_\beta(1) = \alpha_F^u(w + \epsilon) - 1 < 0$}. 

Thus, $\dot{g}_\beta(\cdot)$ will attain negative values as well for some $\beta \in [0,1]$. Since  $\dot{g}_\beta(\cdot)$ is a convex function, the set $D := \{\beta: \dot{g}_\beta(\beta) \leq 0\} = [\tilde{\beta_1}, \tilde{\beta_2}]$ is a connected set, for some $\tilde{\beta_1}, \tilde{\beta_2} \in [0, 1]$ (as in Figure \ref{figure_proof}). Then, $g_\beta(\cdot)$ is decreasing for $\beta \in D$ and increasing otherwise. From this, we infer $g_\beta(\cdot)$ achieves maximum value at $\tilde{\beta_1}$ and minimum value at $\tilde{\beta_2}$, and it equals zero at some $\beta \in D$, i.e., $\beta^*$ in Theorem \ref{thrm_ODE_eta_c}. Further, recall that { $g_\beta(0) > 0$}, {$g_\beta(1) < 0$}, thus,  $g_\beta(\cdot)$ can not have zero in $D^c$. Thus, we have a unique solution for \eqref{Eqn_theta_star_eta_c}. Also clearly, $g_\beta (\beta) > 0$ for all $0< \beta < \beta^*$ and $g_\beta (\beta) < 0$ for all $\beta^* < \beta < 1 $. Henceforth, the proof can be completed as in Lemma \ref{lemma_ASL}. 
\eop
}{}


\noindent \textbf{ Proof of Theorem \ref{thm:monotonic_model1}:}  Consider two systems, with   parameters   $w_1, w_2$    such that $w_1 > w_2$ and the same $b$. Also, the systems start at the same state, i.e., $Z_1(0) = Z_2 (0).$   
 We compare the two systems sample-path wise using   appropriate coupling\footnote{(i.e., chose the random quantities governing the two evolution in such a way that sample path wise comparison is possible)} arguments. Until a wake-up event, both the states remain the same,  hence     can assume the same (say $x$) user  wakes up.  
 At this epoch, $t$, 
   $q_{F1}^u (\omega_{1_t} ) > q_{F2}^u (\omega_{2_t} ) $, as, for $i = 1,2$:
   
   \vspace{-4mm}
{\small
\begin{eqnarray} \label{eq:coupl1}
\vspace{-1mm}
P(T^x_{Fi}=1|\mathcal{G}_t)=q_{Fi}^u = \alpha_i^u \left (\frac{w_i X_i(t)}{X_i(t)+bY_i(t)} +\epsilon\right ).
\end{eqnarray}}
  We now couple the two flags
$T^x_{Fi}$, $i = 1, 2$ as follows:  first generate   flag $T^x_{F1}$   and then set $T^x_{F2} = T^x_{F1}T^x_{F12}$, where flags $T^x_{F1}$, 
$T^x_{F12}$   equal one with the following probabilities: 

\vspace{-4mm}
{\small
\begin{eqnarray}
\label{eq:coupl2}
P(T^x_{F1}=1|\mathcal{G}_t)=q_{F1}^u(\omega_{1_t}),  P(T^x_{F12}=1|\mathcal{G}_t)=\frac{q_{F2}^u(\omega_{2_t})}{q_{F1}^u(\omega_{1_t})}.
\end{eqnarray}}
By virtue of this,  we have that $T^x_{F1} \geq T^x_{F2}$ a.s (as $T^x_{F12} \leq 1)$, i.e.,  in system 1, it is more likely that a user tags a post as fake in comparison with that in system 2. 

If {\small $T^x_{F1}=T^x_{F2}=1$} or {\small $T^x_{F1}=T^x_{F2}=0$}, one can simply couple the offsprings produced by both systems, i.e., set  {\small $\xi_{xx_1} (t) =\xi_{xx_2} (t) =Bin(\mathcal{F},\eta_u)$}, the same realization. 

But if $T^x_{F1}=1 $ and  $ T^x_{F2}=0$, i.e., if in system 1 the user declares the news as fake while in system 2 the user declares the news as real, we couple them as:

\vspace{-3mm}
{\small
\begin{equation}
\label{eq:coup43}
\xi_{xx_1(t)} = \xi_{xy_2(t)} =  Bin(\mathcal{F},\eta_u).
\end{equation}}
Thus we have (as 
 {\small $X_1(t^-) = X_2(t^-), \ Y_1(t^-)= Y_2(t^-)$}):
 
 \vspace{-4mm}
{\small
\begin{eqnarray}
\vspace{-1mm}
        X_1(t^+)  \geq X_2(t^+) &\mbox{ and }&  Y_1(t^+) \leq Y_2(t^+) ~a.s. \label{eq:th4eq2}
\\
    X_1(t^+)+Y_1(t^+)  &=& X_2(t^+)+Y_2(t^+)\  a.s., \mbox{ and hence, } \nonumber \\ 
    \frac{ X_1(t^+) }{X_1(t^+)+Y_1(t^+) } &\ge& \frac{X_2(t^+)}
    {X_2(t^+)+Y_2(t^+)}
    \label{eq:th4eq4}
    \end{eqnarray}}
Hence    again, 
 $q_{F1}^u (\omega_{1_{t^+}} ) > q_{F2}^u (\omega_{2_{t^+}} ) .$ Further because of 
 \eqref{eq:th4eq4}, by appropriate coupling, either the same type wakes up in both systems after time $t$, or  $x$-type  wakes up 
 in system 1 while $y$-type wakes in systems 2.  In either case the probability of the user tagging news   as fake is again bigger in system 1 (as $\alpha_F^u > \alpha_R^u$).
 Using similar coupling logic, we   again have  that (at next wake-up epoch), either both tags  are the same, or $T^x_{F1}=1$ and $T^x_{F2}=0.$  One can progress in the same manner for all time   and the first part  is true. 

\noindent From \eqref{eq:warning}, a decrease in $b$ (for fixed $w$) has   same effects as increase in $w$ (for fixed $b$);  therefore, the result follows. \eop 

\vspace{0.2cm}

\noindent \textbf{Proof of Lemma \ref{lem:opt}:} First observe that solution  $\beta^*$ of \eqref{Eqn_theta_star} (for any $u$) is continuous in $(w, b)$ by Maximum Theorem (\cite[Theorem  9.14,  pp.  235]{sundaram1996first}) and uniqueness of solution of \eqref{Eqn_theta_star}, as $\beta^*$ is the unique minimizer of the following:
\vspace{-1mm}
$$ \left(\beta \left(1 - q_F^u(\beta) + q_R^u(\beta)\right) - q_R^u(\beta)\right)^2.$$
Thus $\Psi_2$ is continuous in $(w, b)$. Consider any point $(w^*,b^*)$ such that $\Psi_2(w^*,b^*) < c$ and consider the following cases:\\
$\bullet$ If {\small $\Psi_2(w^*,b^*) < c < \Psi_2(1,b^*)$}:  by intermediate value theorem (IVT) for the continuous mapping $w \mapsto \psi_2(w, b^*)$, $\exists$ $\varepsilon_w \geq 0$ such that {\small $\Psi_2(w^* + \varepsilon_w,b^*) = c.$}\\
$\bullet$ If not, {\small $\Psi_2(1,b^*) < c$} and hence, 
{\small $\Psi_2(1,0) > c > \Psi_2(1,b^*)$}. By  IVT, $\exists$ an $\varepsilon_b \geq 0$,  
such that {\small $\Psi_2(1,b^* - \varepsilon_b) = ~c.$}

In all, $\exists$ an $\varepsilon_w, \varepsilon_b \geq 0$ such that $\Psi_2(w^* + \varepsilon_w, b^* - \varepsilon_b) = c$. By Theorem \ref{thm:monotonic_model1}, $\Psi_1(w^* + \varepsilon_w, b^* - \varepsilon_b) \leq \Psi_1 (w^*, b^*)$. Hence the lemma follows. \eop


\Arxivtwo{

}{
\bibliographystyle{IEEEtran}
\bibliography{my_ref}
}

\end{document}

%% file: 1.pdf_tex
\begingroup%
  \makeatletter%
  \providecommand\color[2][]{%
    \errmessage{(Inkscape) Color is used for the text in Inkscape, but the package 'color.sty' is not loaded}%
    \renewcommand\color[2][]{}%
  }%
  \providecommand\transparent[1]{%
    \errmessage{(Inkscape) Transparency is used (non-zero) for the text in Inkscape, but the package 'transparent.sty' is not loaded}%
    \renewcommand\transparent[1]{}%
  }%
  \providecommand\rotatebox[2]{#2}%
  \newcommand*\fsize{\dimexpr\f@size pt\relax}%
  \newcommand*\lineheight[1]{\fontsize{\fsize}{#1\fsize}\selectfont}%
  \ifx\svgwidth\undefined%
    \setlength{\unitlength}{765.35433071bp}%
    \ifx\svgscale\undefined%
      \relax%
    \else%
      \setlength{\unitlength}{\unitlength * \real{\svgscale}}%
    \fi%
  \else%
    \setlength{\unitlength}{\svgwidth}%
  \fi%
  \global\let\svgwidth\undefined%
  \global\let\svgscale\undefined%
  \makeatother%
  \begin{picture}(1,0.96296296)%
    \lineheight{1}%
    \setlength\tabcolsep{0pt}%
    \put(0,0){\includegraphics[width=\unitlength,page=1]{1.pdf}}%
    \put(0.33468662,0.37762101){\color[rgb]{0,0,0}\makebox(0,0)[lt]{\lineheight{1.25}\smash{\begin{tabular}[t]{l}$\tilde{\beta_1}$\end{tabular}}}}%
    \put(0.74552607,0.3756615){\color[rgb]{0,0,0}\makebox(0,0)[lt]{\lineheight{1.25}\smash{\begin{tabular}[t]{l}$\tilde{\beta_2}$\end{tabular}}}}%
    \put(0.88642945,0.38363331){\color[rgb]{0,0,0}\makebox(0,0)[lt]{\lineheight{1.25}\smash{\begin{tabular}[t]{l}1\end{tabular}}}}%
    \put(0.18541049,0.40169904){\color[rgb]{0,0,0}\makebox(0,0)[lt]{\lineheight{1.25}\smash{\begin{tabular}[t]{l}0\end{tabular}}}}%
    \put(0.96625005,0.43502632){\color[rgb]{0,0,0}\makebox(0,0)[lt]{\lineheight{1.25}\smash{\begin{tabular}[t]{l}$\beta$\end{tabular}}}}%
    \put(0.79400556,0.8565354){\color[rgb]{0,0,0}\makebox(0,0)[lt]{\lineheight{1.25}\smash{\begin{tabular}[t]{l}$\dot{g}_\beta(\beta)$\end{tabular}}}}%
    \put(0,0){\includegraphics[width=\unitlength,page=2]{1.pdf}}%
    \put(0.79260514,0.77448297){\color[rgb]{0,0,0}\makebox(0,0)[lt]{\lineheight{1.25}\smash{\begin{tabular}[t]{l}$g_\beta(\beta)$\end{tabular}}}}%
    \put(0,0){\includegraphics[width=\unitlength,page=3]{1.pdf}}%
  \end{picture}%
\endgroup%

%% file: 4.pdf_tex
\begingroup%
  \makeatletter%
  \providecommand\color[2][]{%
    \errmessage{(Inkscape) Color is used for the text in Inkscape, but the package 'color.sty' is not loaded}%
    \renewcommand\color[2][]{}%
  }%
  \providecommand\transparent[1]{%
    \errmessage{(Inkscape) Transparency is used (non-zero) for the text in Inkscape, but the package 'transparent.sty' is not loaded}%
    \renewcommand\transparent[1]{}%
  }%
  \providecommand\rotatebox[2]{#2}%
  \newcommand*\fsize{\dimexpr\f@size pt\relax}%
  \newcommand*\lineheight[1]{\fontsize{\fsize}{#1\fsize}\selectfont}%
  \ifx\svgwidth\undefined%
    \setlength{\unitlength}{765.35433071bp}%
    \ifx\svgscale\undefined%
      \relax%
    \else%
      \setlength{\unitlength}{\unitlength * \real{\svgscale}}%
    \fi%
  \else%
    \setlength{\unitlength}{\svgwidth}%
  \fi%
  \global\let\svgwidth\undefined%
  \global\let\svgscale\undefined%
  \makeatother%
  \begin{picture}(1,0.96296296)%
    \lineheight{1}%
    \setlength\tabcolsep{0pt}%
    \put(0,0){\includegraphics[width=\unitlength,page=1]{4.pdf}}%
    \put(0.71324107,0.39768304){\color[rgb]{0,0,0}\makebox(0,0)[lt]{\lineheight{1.25}\smash{\begin{tabular}[t]{l}$\tilde{\beta_2}$\end{tabular}}}}%
    \put(0.81427861,0.41501499){\color[rgb]{0,0,0}\makebox(0,0)[lt]{\lineheight{1.25}\smash{\begin{tabular}[t]{l}1\end{tabular}}}}%
    \put(0.9058583,0.45464896){\color[rgb]{0,0,0}\makebox(0,0)[lt]{\lineheight{1.25}\smash{\begin{tabular}[t]{l}$\beta$\end{tabular}}}}%
    \put(0,0){\includegraphics[width=\unitlength,page=2]{4.pdf}}%
    \put(0.77588235,0.85140288){\color[rgb]{0,0,0}\makebox(0,0)[lt]{\lineheight{1.25}\smash{\begin{tabular}[t]{l}$\dot{g}_\beta(\beta)$\end{tabular}}}}%
    \put(0,0){\includegraphics[width=\unitlength,page=3]{4.pdf}}%
    \put(0.77448194,0.76935043){\color[rgb]{0,0,0}\makebox(0,0)[lt]{\lineheight{1.25}\smash{\begin{tabular}[t]{l}$g_\beta(\beta)$\end{tabular}}}}%
    \put(0,0){\includegraphics[width=\unitlength,page=4]{4.pdf}}%
    \put(0.37631601,0.15153453){\makebox(0,0)[lt]{\lineheight{1.25}\smash{\begin{tabular}[t]{l}      ($\tilde{\beta_1}= 0$)\end{tabular}}}}%
    \put(0.11597672,0.41308967){\makebox(0,0)[lt]{\lineheight{1.25}\smash{\begin{tabular}[t]{l}0\end{tabular}}}}%
  \end{picture}%
\endgroup%

%% file: drawing.pdf_tex
\begingroup%
  \makeatletter%
  \providecommand\color[2][]{%
    \errmessage{(Inkscape) Color is used for the text in Inkscape, but the package 'color.sty' is not loaded}%
    \renewcommand\color[2][]{}%
  }%
  \providecommand\transparent[1]{%
    \errmessage{(Inkscape) Transparency is used (non-zero) for the text in Inkscape, but the package 'transparent.sty' is not loaded}%
    \renewcommand\transparent[1]{}%
  }%
  \providecommand\rotatebox[2]{#2}%
  \newcommand*\fsize{\dimexpr\f@size pt\relax}%
  \newcommand*\lineheight[1]{\fontsize{\fsize}{#1\fsize}\selectfont}%
  \ifx\svgwidth\undefined%
    \setlength{\unitlength}{765.35433071bp}%
    \ifx\svgscale\undefined%
      \relax%
    \else%
      \setlength{\unitlength}{\unitlength * \real{\svgscale}}%
    \fi%
  \else%
    \setlength{\unitlength}{\svgwidth}%
  \fi%
  \global\let\svgwidth\undefined%
  \global\let\svgscale\undefined%
  \makeatother%
  \begin{picture}(1,0.96296296)%
    \lineheight{1}%
    \setlength\tabcolsep{0pt}%
    \put(0,0){\includegraphics[width=\unitlength,page=1]{drawing.pdf}}%
    \put(0.80329093,0.41779175){\color[rgb]{0,0,0}\makebox(0,0)[lt]{\lineheight{1.25}\smash{\begin{tabular}[t]{l}1\end{tabular}}}}%
    \put(0.11011142,0.41037942){\color[rgb]{0,0,0}\makebox(0,0)[lt]{\lineheight{1.25}\smash{\begin{tabular}[t]{l}0\end{tabular}}}}%
    \put(0.89291077,0.44958626){\color[rgb]{0,0,0}\makebox(0,0)[lt]{\lineheight{1.25}\smash{\begin{tabular}[t]{l}$\beta$\end{tabular}}}}%
    \put(0,0){\includegraphics[width=\unitlength,page=2]{drawing.pdf}}%
    \put(0.74502558,0.84963082){\color[rgb]{0,0,0}\makebox(0,0)[lt]{\lineheight{1.25}\smash{\begin{tabular}[t]{l}$\dot{g}_\beta(\beta)$\end{tabular}}}}%
    \put(0,0){\includegraphics[width=\unitlength,page=3]{drawing.pdf}}%
    \put(0.74362516,0.76757837){\color[rgb]{0,0,0}\makebox(0,0)[lt]{\lineheight{1.25}\smash{\begin{tabular}[t]{l}$g_\beta(\beta)$\end{tabular}}}}%
    \put(0,0){\includegraphics[width=\unitlength,page=4]{drawing.pdf}}%
    \put(0.51614391,0.08778833){\makebox(0,0)[lt]{\lineheight{1.25}\smash{\begin{tabular}[t]{l}($\tilde{\beta_1} = 0$, $\tilde{\beta_2} = 1$)\end{tabular}}}}%
  \end{picture}%
\endgroup%

%% file: FakeNewsAfterReviews.bbl
\begin{thebibliography}{} 
\bibitem{lazer2018science} Lazer, David MJ, et al. "The science of fake news." Science 359.6380 (2018): 1094-1096.
\bibitem{Allcott} Allcott, Hunt, and Matthew Gentzkow. "Social media and fake news in the 2016 election." Journal of economic perspectives 31.2 (2017): 211-36.
\bibitem{Kogan}Kogan, Shimon, Tobias J. Moskowitz, and Marina Niessner. "Fake news: Evidence from financial markets." Available at SSRN 3237763 (2019).
\bibitem{Vosoughi1146}Vosoughi, Soroush, Deb Roy, and Sinan Aral. "The spread of true and false news online." Science 359.6380 (2018): 1146-1151.
\bibitem{zhao2018fake} Zhao, Zilong, et al. "Fake news propagates differently from real news even at early stages of spreading." EPJ Data Science 9.1 (2020): 7.
\bibitem{viralBranching}Van der Lans, Ralf, et al. "A viral branching model for predicting the spread of electronic word of mouth." Marketing Science 29.2 (2010): 348-365.
\bibitem{ranbir} Dhounchak, Ranbir, Veeraruna Kavitha, and Eitan Altman. "A viral timeline branching process to study a social network." 2017 29th International Teletraffic Congress (ITC 29). Vol. 3. IEEE, 2017.
\bibitem{zhou2018fake}Zhou, Xinyi, and Reza Zafarani. "Fake news: A survey of research, detection methods, and opportunities." arXiv preprint arXiv:1812.00315 (2018).
\bibitem{landemore2012collective}Landemore, Hélène, and Jon Elster, eds. Collective wisdom: Principles and mechanisms. Cambridge University Press, 2012.
\bibitem{gonzalez2004multitype}González, Miguel, Rodrigo Martínez, and Manuel Mota. "Multitype population size-dependent branching processes with dependent offspring." Statistics \& probability letters 70.2 (2004): 145-154.
\bibitem{snapnets}Leskovec, Jure, and Andrej Krevl. "SNAP Datasets: Stanford large network dataset collection." (2014).
\bibitem{athreya}Williamson, John A. "KB Athreya, PE Ney, Branching Processes." The Annals of Probability 2.5 (1974): 966-968.
\bibitem{kushner2003stochastic}Kushner, Harold, and G. George Yin. Stochastic approximation and recursive algorithms and applications. Vol. 35. Springer Science \& Business Media, 2003.
\bibitem{sundaram1996first} Sundaram, Rangarajan K. A first course in optimization theory. Cambridge university press, 1996.
\bibitem{agarwal2021new} Agarwal, Khushboo, and Veeraruna Kavitha. "New results in Branching processes using Stochastic Approximation." submitted and preprint available as arXiv preprint arXiv:2111.14527 (2021).
\bibitem{d2019spreading} D O’Brien, Joseph, Ioannis K. Dassios, and James P. Gleeson. "Spreading of memes on multiplex networks." New Journal of Physics 21.2 (2019): 025001.
\bibitem{yagan2013conjoining} Yagan, Osman, et al. "Conjoining speeds up information diffusion in overlaying social-physical networks." IEEE Journal on Selected Areas in Communications 31.6 (2013): 1038-1048.
\bibitem{arxiv} Kapsikar, Suyog, et al. "Controlling fake news by tagging: A branching process analysis." arXiv preprint arXiv:2009.02275 (2020).
\bibitem{klebaner1989geometric} Klebaner, Fima C. "Geometric growth in near-supercritical population size dependent multitype Galton-Watson processes." The Annals of Probability 17.4 (1989): 1466-1477.
\end{thebibliography}
